%% file: main.tex
\pdfoutput=1
\documentclass[sigconf,nonacm]{acmart}
\usepackage[capitalize,noabbrev]{cleveref}
\usepackage{amsmath}  
\newcommand{\model}[1]{\mathrm{#1}}
\usepackage{wrapfig}

\setcopyright{none}
\AtBeginDocument{%
  }

\begin{document}

\title{TBGRecall: A Generative Retrieval Model for E-commerce Recommendation Scenarios}



\author{Zida Liang}
\authornote{Both authors contributed equally to this research.}
\authornote{Work done during internship at Alibaba.}
\affiliation{%
  \institution{Shanghai Jiaotong University}
  \city{Shanghai}
  \country{China}}
\email{greek-guardian@sjtu.edu.cn}

\author{Changfa Wu}
\authornotemark[1]
\affiliation{%
  \institution{Alibaba Inc.}
  \city{Beijing}
  \country{China}}
\email{wuchangfa.wcf@alibaba-inc.com}

\author{Dunxian Huang}
\affiliation{%
  \institution{Alibaba Inc.}
  \city{Hangzhou}
  \state{Zhejiang}
  \country{China}}
\email{dunxian.hdx@alibaba-inc.com}

\author{Weiqiang Sun}
\affiliation{%
  \institution{Shanghai Jiaotong University}
  \city{Shanghai}
  \country{China}}
\email{sunwq@sjtu.edu.cn}

\author{Ziyang Wang}
\affiliation{%
  \institution{Alibaba Inc.}
  \city{Hangzhou}
  \state{Zhejiang}
  \country{China}}
\email{shanyi.wzy@alibaba-inc.com}

\author{Yuliang Yan}
\affiliation{%
  \institution{Alibaba Inc.}
  \city{Hangzhou}
  \state{Zhejiang}
  \country{China}}
\email{yuliang.yyl@alibaba-inc.com}
\author{Jian Wu }
\affiliation{%
  \institution{Alibaba Inc.}
  \city{Beijing}
  \country{China}}
\email{joshuawu.wujian@alibaba-inc.com}

\author{Yuning Jiang }
\affiliation{
  \institution{Alibaba Inc.}
  \city{Beijing}
  \country{China}
}
\email{mengzhu.jyn@alibaba-inc.com}
\author{Bo Zheng }
\affiliation{
  \institution{Alibaba Inc.}
  \city{Beijing}
  \country{China}
}
\email{bozheng@alibaba-inc.com}
\author{Ke Chen }
\affiliation{
  \institution{Alibaba Inc.}
  \city{Hangzhou}
  \state{Zhejiang}
  \country{China}
}
\email{ck325090@alibaba-inc.com}
\author{Silu Zhou }
\affiliation{
  \institution{Alibaba Inc.}
  \city{Hangzhou}
  \state{Zhejiang}
  \country{China}
}
\email{silu.zsl@alibaba-inc.com}
\author{Yu Zhang }
\affiliation{
  \institution{Alibaba Inc.}
  \city{Hangzhou}
  \state{Zhejiang}
  \country{China}
}
\email{zy429782@alibaba-inc.com}

\renewcommand{\shortauthors}{Zida Liang et al.}

\begin{abstract}
Recommendation systems are essential tools in modern e-commerce, facilitating personalized user experiences by suggesting relevant products. Recent advancements in generative models have demonstrated potential in enhancing recommendation systems; however, these models often exhibit limitations in optimizing retrieval tasks, primarily due to their reliance on autoregressive generation mechanisms. Conventional approaches introduce sequential dependencies that impede efficient retrieval, as they are inherently unsuitable for generating multiple items without positional constraints within a single request session. To address these limitations, we propose TBGRecall, a framework integrating Next Session Prediction (NSP), designed to enhance generative retrieval models for e-commerce applications. Our framework reformulation involves partitioning input samples into multi-session sequences, where each sequence comprises a session token followed by a set of item tokens, and then further incorporate multiple optimizations tailored to the generative task in retrieval scenarios. In terms of training methodology, our pipeline integrates limited historical data pre-training with stochastic partial incremental training, significantly improving training efficiency and emphasizing the superiority of data recency over sheer data volume. Our extensive experiments, conducted on public benchmarks alongside a large-scale industrial dataset from TaoBao, show TBGRecall outperforms the state-of-the-art recommendation methods, and exhibits a clear scaling law trend. Ultimately, NSP represents a significant advancement in the effectiveness of generative recommendation systems for e-commerce applications.
\end{abstract}

\ccsdesc[500]{Information systems~Recommender systems}

\keywords{Recommendation System, Retrieval stage, Next Session Prediction}


\maketitle

\input{introduction}

\input{related_work}

\input{method}

\input{Engineering_level_details}

\input{experiment}

\input{conclusion}

\clearpage



\bibliographystyle{ACM-Reference-Format}
\bibliography{main}



\end{document}

%% file: introduction.tex
\section{Introduction}
Recommendation systems~\cite{isinkaye2015recommendation,ko2022survey,gao2023survey} are critical to e-commerce~\cite{pfadler2020billion} and social platforms~\cite{fan2019graph}. The retrieval stage~\cite{huang2024comprehensive}, the pipeline’s foundational step, generates a large candidate item set (e.g., thousands) for subsequent ranking modules, including rough ranking and hybrid ranking. Notably, this process resembles a generative task, making it amenable to modern generative modeling paradigms.

Generative neural models, which excel in NLP~\cite{chenlonglora,touvron2023llama} and CV~\cite{radford2021learning,liang2024survey}, have recently gained traction in recommendation systems. A prominent direction involves leveraging large language models (LLMs): some works use LLMs for feature enhancement in retrieval tasks \cite{googleStar, metaEmbsum}, while others explore direct LLM-based item generation \cite{kuaishouOnerec}. However, challenges like semantic alignment limit their efficacy. Alternative approaches adapt LLM paradigms for recommendations, such as Tiger's\cite{googleTiger} RQ-VAE-based semantic ID framework, later extended in \cite{tencentLCRec, baiduCOBRA}. Other studies propose sparse ID generation via models like HSTU \cite{metaHSTU} and E4SRec \cite{thuE4SRec}. And some other research like sessionrec \cite{huang2025sessionrec} uses new methods for generative recommend.

Despite recent progress, existing generative models lack task-specific optimization for retrieval pipelines. In retrieval, systems return unordered candidate sets per session—a recommendation request with no intra-session item dependencies. Conventional token-wise autoregressive models, which enforce sequential item dependencies, violate retrieval requirements: user interactions occur after item exposure, creating causality only across sessions. Thus, items within a session should not influence each other but inform future sessions.

To address this, we propose \textbf{Taobao Generative Recall} (TBGRecall), a session-wise autoregressive model where item generation is independent. Its core innovation is Next Session Prediction (NSP), a generation paradigm optimized for retrieval. By reengineering sample structure and loss functions, NSP outperforms classical Next Token Prediction (NTP). Built on HSTU \cite{metaHSTU}, our framework partitions user sequences into multi-session segments, each starting with a {session token} followed by {item tokens}. During training, contrastive loss between context tokens and positive/negative item samples replaces regressive loss. At inference, the session token guides Approximate Nearest Neighbor (ANN) search for retrieval. This design resolves autoregressive dependency conflicts and aligns with retrieval infrastructure.

Further optimizations enhance NSP: multi-session prediction (MSP) and token-specific network (TSN) improve performance, while stochastic partial incremental training reduces computational costs for large-scale models. Our framework exhibits scaling law trends and surpasses baselines. NSP represents a pivotal advancement in generative retrieval, bridging the gap between autoregressive modeling and practical recommendation pipelines.

\begin{itemize}
\item We introduced Next Session Prediction for non-autoregressive generation, along with a set of optimizations such as MSP, outperforming Next Token Prediction in recommendation retrieval tasks.
\item We proposed stochastic partial incremental training, which significantly improves training efficiency without compromising model performance, highlighting the superiority of data recency over sheer data volume.
\item TBGRecall outperforms SOTA methods in offline experiments. We have implemented it with nearline on Taobao’s HomePage "Guess You Like" Scenarios, resulting in a significant online transaction amount increase of 2.16\%.
\end{itemize}

%% file: related_work.tex
\section{Related Work}

\begin{figure*}[ht]
\begin{center}
\includegraphics[width=\textwidth]{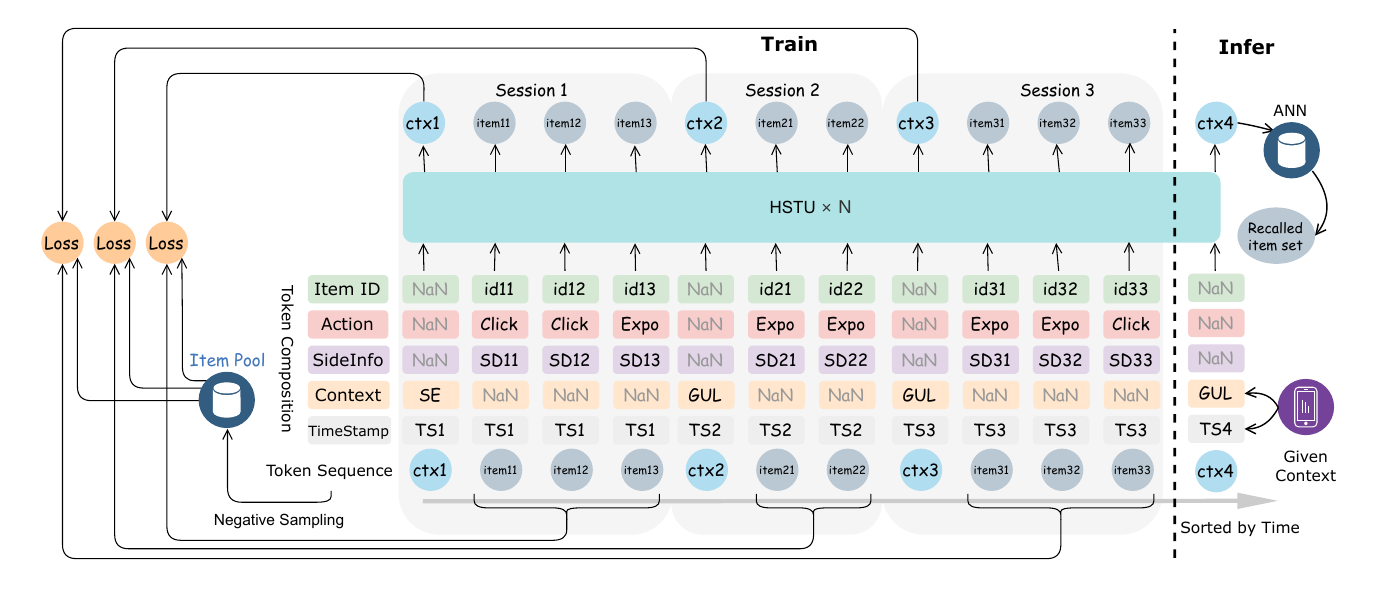}
\vskip -0.15in
\caption{TBGRecall structure. The figure illustrates the overall architecture of the model, the construction of the sequence, the composition of the tokens, and the computation of the loss. Since the two types of tokens correspond to different inputs, some inputs will be defined as NaN, and they also have a corresponding embedding.}
\label{tbgrecall}
\end{center}
\vskip -0.15in
\end{figure*}

\subsection{Traditional Retrieval Methods}
In recommendation systems, the retrieval phase serves as a critical first step to filter vast item corpora into a relevant subset for users, based on historical interactions and behavioral patterns. Traditional approaches include content-based filtering—which leverages item attributes (e.g., keywords, categories) to recommend similar items—and collaborative filtering, which exploits user-item interaction data to identify preferences through community behavior \cite{CF}. Notably, item-based collaborative filtering \cite{CF} and matrix factorization (MF) \cite{MF} demonstrate superior performance by modeling latent relationships and integrating implicit feedback, temporal dynamics, and confidence signals. Large-scale applications further advance these ideas: the Swing algorithm \cite{SWING} constructs product graphs to capture item substitutability and complementarity in e-commerce environments like Taobao. Recent breakthroughs, however, are driven by deep learning-based embedding models, particularly two-tower architectures deployed at YouTube \cite{youtubednn, TTM2}, which enhance candidate generation through unbiased representation learning from streaming data while addressing sampling biases. These advancements highlight the transformative impact of deep learning on scalability and precision in large-scale recommendation systems.
\subsection{Generative Models for Recommendation}

Recent advances in LLMs enable semantic feature augmentation and generative modeling for recommendation systems. Google proposes a training-free framework integrating LLM embeddings via two-stage retrieval-ranking\cite{googleStar}. Meta’s EmbSum\cite{metaEmbsum} uses a pretrained encoder-decoder with poly-attention to generate UPE/CPE, refined by LLM-supervised summarization. Bytedance’s HLLM\cite{bytedanceHLLM} decouples content/behavior analysis via Item/User LLMs, improving sequential modeling accuracy.

In sequential recommendation, generative models—particularly sequence-to-sequence architectures—have emerged as a transformative paradigm by framing next-item prediction as autoregressive generation. Early Transformer-based approaches like SASRec demonstrated the efficacy of self-attention in modeling masked user sequences. Building on this, google\cite{googleTiger} introduces a content-quantized framework encoding item semantics into discrete semantic-ID tuples, with a decoder generating these tokens to enhance cold-start robustness and diversity control. Tencent\cite{tencentLCRec} bridges language and collaborative signals via vector quantization and alignment tuning, better integrating recommendation-specific features into LLMs. Kuaishou\cite{kuaishouOnerec} further unifies retrieval and ranking within a single model through sparse MoE, session-wise generation, and DPO-based alignment. Recent works extend these ideas: Meta\cite{metaHSTU} treats recommendation as sequential transduction, enabling efficient scaling in high-cardinality environments\cite{zhang2023scalinglawlargesequential}, while others\cite{thuE4SRec} combines sequential pretraining with LLM instruction tuning by embedding item IDs into prompts, enabling both accurate retrieval and interpretable recommendations.

%% file: method.tex
\section{Methodology}

\subsection{Problem Formulation}

Taobao is one of the leading e-commerce platforms in the industry, providing consumers and merchants with a secure and convenient channel for buying and selling goods. The majority of Taobao’s business traffic comes from the mobile Taobao app, which includes various scenarios such as homepage product feed (Guess You Like, abbreviated as GUL), inshop (IS), and search (SE), among others.

The model proposed in this paper, named TBGRecall , focuses primarily on item retrieval in the "Guess You Like" (GUL) scenario within product recommendation systems. Our core contribution lies in the introduction of a novel method called \textbf{Next Session Prediction} to obtain a session-level embedding that represents the user's current intent. Based on this embedding, we perform Approximate Nearest Neighbor (ANN) search to retrieve candidate items.

\begin{equation}
    \mathbf{h}'_{c^{(K+1)}} = \mathcal{M}_{\model{TBGRecall}}(\mathcal{Q})
    \label{model_equation}
\end{equation}

In \cref{model_equation}, $(K+1)$ denotes the $(K+1)$-th session of a certain user behaviors, and $\mathbf{h}'_{c^{(K+1)}}$ represents the output at the corresponding position of the context token. 
In TBGRecall, given user interaction sequence $\mathcal{Q}$, $\mathbf{h}'_{c^{(K+1)}}$ serves as the final user latent vector transmitted to the ANN.

\subsection{Sample Construction}
\label{Sequence_Construction_subsection}

In the homepage recommendation setting, each time a user opens the homepage or scrolls down the feed, a recommendation request is triggered on the client side, which is referred to each as a \emph{session}. Upon receiving a session request, the backend recommendation system returns a curated set of items to be displayed to the user. The user can then interact with these items, generating user actions. Other scenarios follow the same request-response pattern. 

In our task, samples are organized at the user level, with each row representing their most recent $K$ session-based interactions aggregated chronologically for intent modeling. For each user, recent sessions are aggregated chronologically into a unified behavioral sequence $\mathcal{Q}$, forming the foundation for modeling intent and purchase dynamics. Content from other verticals such as advertising and livestreaming is excluded, which results in varying numbers of items per session. For each item in a session, we also record associated side information—such as first- and second-level category IDs, seller ID, price, and the time of recommendation—as part of the behavior sequence. User behaviors, including clicks, are also recorded in the sequence. 
As illustrated in \cref{tbgrecall}, we segment the nearest behavior sequence of user into multiple sessions, where $K$ is the number of sessions and each session starts with a context token $c^{(k)}$ followed by $N_k$ item tokens, where $c^{(k)}$ represents the session scenario, and $i^{(k)}_m$ denotes the $m$-th item in session $k$. The model input sequence $\mathcal{Q}$ is:

\begin{equation}
    \mathcal{Q} = \left[c^{(1)}, i^{(1)}_1, i^{(1)}_2, i^{(1)}_3, c^{(2)},\dots,c^{(K)}, i^{(K)}_1, i^{(K)}_2, c^{(K+1)} \right],
\end{equation}

where $c^{(K+1)}$ is the context token of the newly arrived session. Each element in the sequence $\mathcal{Q}$ corresponds to a token in the model, which is obtained by computing the embeddings of its associated information and summing them together. Specifically, each token's hidden vector $\mathbf{h}$ is composed of four components:

\begin{equation}
    \mathbf{h} = \mathbf{e}_{\text{id}} + \mathbf{e}_{\text{act}} + \mathbf{e}_{\text{side}} + \mathbf{e}_{\text{ctx}},
    \label{hidden_embedding}
\end{equation}

where $\mathbf{e}_{\text{id}}$ is the item ID embedding, $\mathbf{e}_{\text{act}}$ is the action embedding representing the type of user interaction (click or exposure), $\mathbf{e}_{\text{side}}$ is the side information embedding capturing auxiliary item attributes (e.g., category, price, seller), and $\mathbf{e}_{\text{ctx}}$ is the embedding of scenarios.


\subsection{Next Session Prediction Framework}

Most existing generative retrieval methods adopt an autoregressive generation paradigm, which assumes a sequential dependency among items within a session. However, in real-world recommendation systems, item order is often semantically meaningless: all items are delivered to the user simultaneously with no inherent causality or temporal relationship among them. This discrepancy leads to unnecessary dependencies among items in autoregressive generation, which may harm both modeling accuracy and inference efficiency. To address this issue, we propose a \textbf{session-wise autoregressive} approach via NSP. This method eliminates the need for iterative item generation during inference, thereby reducing online computation overhead while also breaking the undesired order dependency among items. Moreover, it aligns naturally with the existing infrastructure of item retrieval systems.


\begin{figure}[ht]
\begin{center}
\vskip -0.15in
\centerline{\includegraphics[width=125pt]{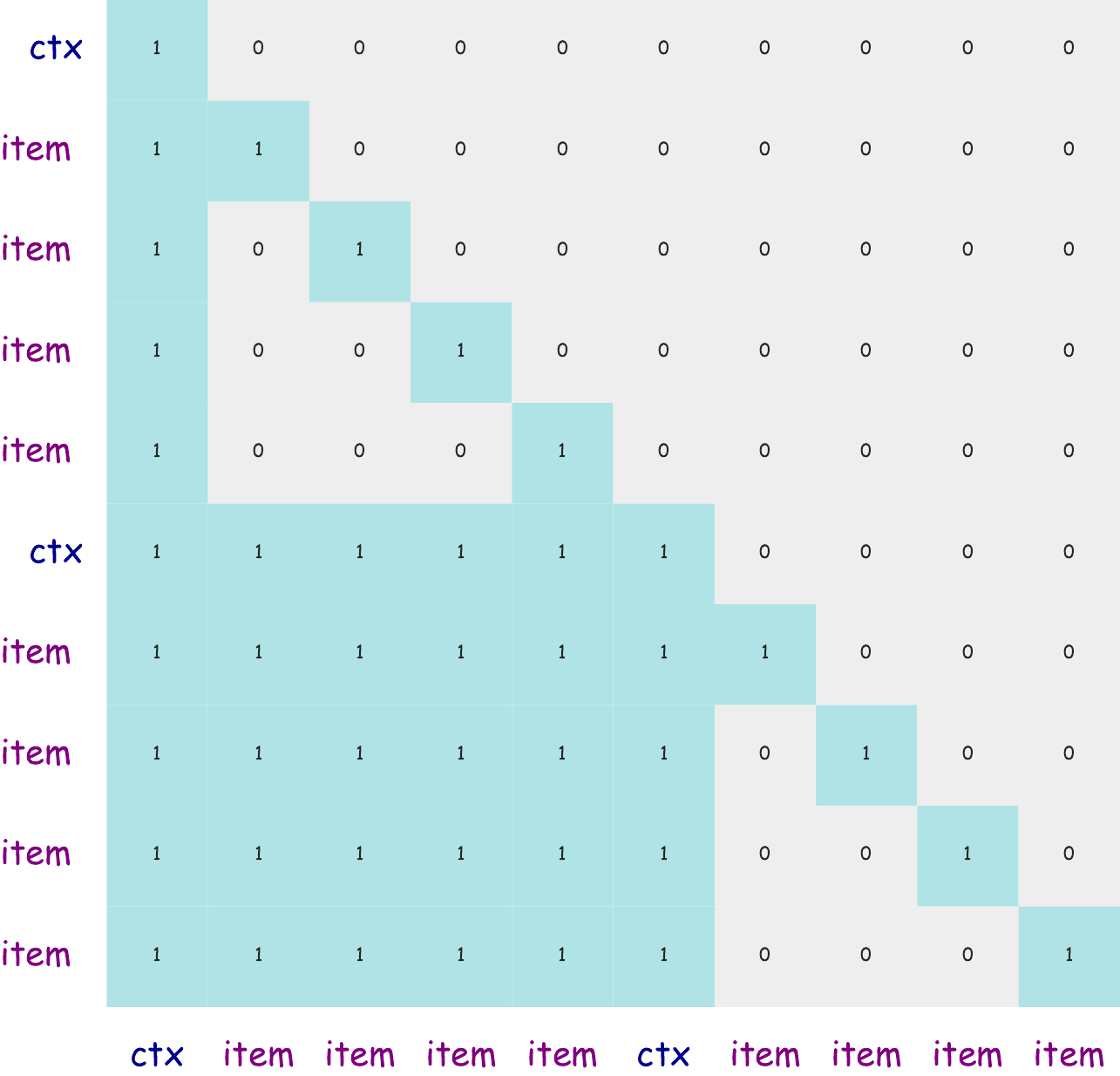}    }
\vskip -0.10in
\caption{Session Mask. This mask is built upon the causal mask and further renders the interactions between items within the same session invisible.}
\vskip -0.2in
\label{session_mask}
\end{center}
\end{figure}


\paragraph{Session Mask} Causal mask is used in decoder-only architecture to restrict attention computation. However, in NSP (Next Session Prediction), we do not want items within the same session to interact with each other. Therefore, we introduce a \textbf{Session Mask} based on the causal mask. 
Besides, instead of using sequential indices for positional encoding, TBGRecall utilizes session indices as seesion-wise rope (sw-rope), where all tokens within the same session share a common positional encoding, to model user session-wise relative positional dependencies.
\paragraph{TSN} To address the semantic and distributional discrepancies between context and item tokens, we introduce a \textbf{Token-Specific Network} (TSN), which applies dedicated linear transformation layers at both the embedding and initial Transformer stages. This architecture eliminates performance degradation caused by shared projections while maintaining inference efficiency—no additional computational overhead is incurred, as TSN merely substitutes a unified projection with two token-type-specific counterparts. 


\paragraph{MSP and MoE}
Given TBGRecall’s alignment with mainstream generative models, we integrate two established LLM techniques—Multi-Token Prediction (MTP) and Mixture-of-Experts (MoE)—to enhance performance. We implement \textbf{Multi-Session Prediction} (MSP) along the session dimension for context tokens, mirroring the MTP paradigm \cite{deepseekv3, metamtp}. This introduces extended training signals, enabling explicit modeling of long-range user behavior dependencies and transitional relationships across distant contextual scenarios. For the feed-forward module, we adopt DeepSeekMoE with auxiliary-loss-free load balancing, which harnesses specialized expert networks to acquire richer, task-specific knowledge at comparable inference FLOPS, thereby improving overall model efficacy.


\subsection{Overall Loss}

Our overall loss function consists of two components: a contrastive loss $\mathcal{L}_{NCE}$ for autoregressive training, and a cascade loss that emphasizes high-value samples, which includes both $\mathcal{L}_{click}$ and $\mathcal{L}_{pay}$. Our proposed loss function is fundamentally derived from Noise Contrastive Estimation (NCE), with tailored modifications designed to better suit the characteristics of recommendation tasks and to enhance the effectiveness of contrastive learning in high-cardinality item spaces. In addition, due to the highly imbalanced distribution of session counts across different scenarios, directly optimizing a unified loss without scenario distinction may lead to performance degradation. To address this issue, we propose Multi-Scene Normalization, where losses are computed independently for each scenario. These per-scenario losses are then normalized and aggregated to ensure balanced optimization across all scenarios. Based on the above, the formulation can be summarized as follows:

\begin{equation}
  \begin{aligned}
    \mathcal{L}_{NSP} = \sum_{s \in \mathcal{S}}\frac{1}{N_s}\left(\mathcal{L}_{NCE}^{(s)} + \mathcal{L}_{click}^{(s)} + \mathcal{L}_{pay}^{(s)}\right)
  \end{aligned}
\end{equation}

Here, $\mathcal{S}$ denotes the set of all sessions, and $N_s$ is the number of sessions under a particular scenario. 

The contrastive loss for a given session $s$ is defined as:

\begin{equation}
  \begin{aligned}
    \mathcal{L}_{NCE}^{(s)} = \frac{1}{|\mathcal{I}^{(s)}|} \sum_{i \in \mathcal{I}^{(s)}} -\log \left( \frac{p^{(s)}_i}{p^{(s)}_i + \sum_{j \in \mathcal{N}^{(s)}} p^{(s)}_j} \right)
  \end{aligned}
\end{equation}

$\mathcal{I}^{(s)}$ represents the set of items recommended to the user within session $s$, serving as positive samples, while $\mathcal{N}^{(s)}$ denotes a set of negative samples randomly drawn from the item pool. The term $p^{(s)}$ corresponds to the softmax-normalized inner product between the context token and the sample tokens. The goal of $\mathcal{L}_{NCE}$ is to bring the context token closer to the positive sample tokens while pushing it away from the negative ones.

To better capture user engagement and transaction value, we incorporate two additional cascade loss terms:

\begin{equation}
  \begin{aligned}
    \mathcal{L}_{click}^{(s)} = \frac{1}{|\mathcal{C}^{(s)}|} \sum_{i \in \mathcal{C}^{(s)}} -\log \left( \frac{p^{(s)}_i}{p^{(s)}_i + \sum_{j \in (\mathcal{I}^{(s)}\setminus\mathcal{C}^{(s)})} p^{(s)}_j} \right)
  \end{aligned}
\end{equation}

\begin{equation}
  \begin{aligned}
    \mathcal{L}_{pay}^{(s)} = \frac{1}{|\mathcal{P}^{(s)}|} \sum_{i \in \mathcal{P}^{(s)}} -\log \left( \frac{p^{(s)}_i}{p^{(s)}_i + \sum_{j \in (\mathcal{C}^{(s)}\setminus\mathcal{P}^{(s)})} p^{(s)}_j} \right)
  \end{aligned}
\end{equation}

Here, $\mathcal{C}^{(s)}$ and $\mathcal{P}^{(s)}$ denote the sets of clicked and purchased items within session $s$, respectively. The losses $\mathcal{L}_{click}$ and $\mathcal{L}_{pay}$ are designed to assign greater importance to samples associated with user clicks, encouraging the model to focus more on high-value interactions.

\subsection{Partial Incremental Training}

For TBGRecall, due to the massive number of model parameters and the scale of user data, training on the newly added data from a single day can take up to five days. This results in a deployment delay, which causes a significant degradation in performance—an outcome that is unacceptable in practice.

\begin{figure}[ht]
\begin{center}
\vskip -0.15in
\centerline{\includegraphics[width=\columnwidth]{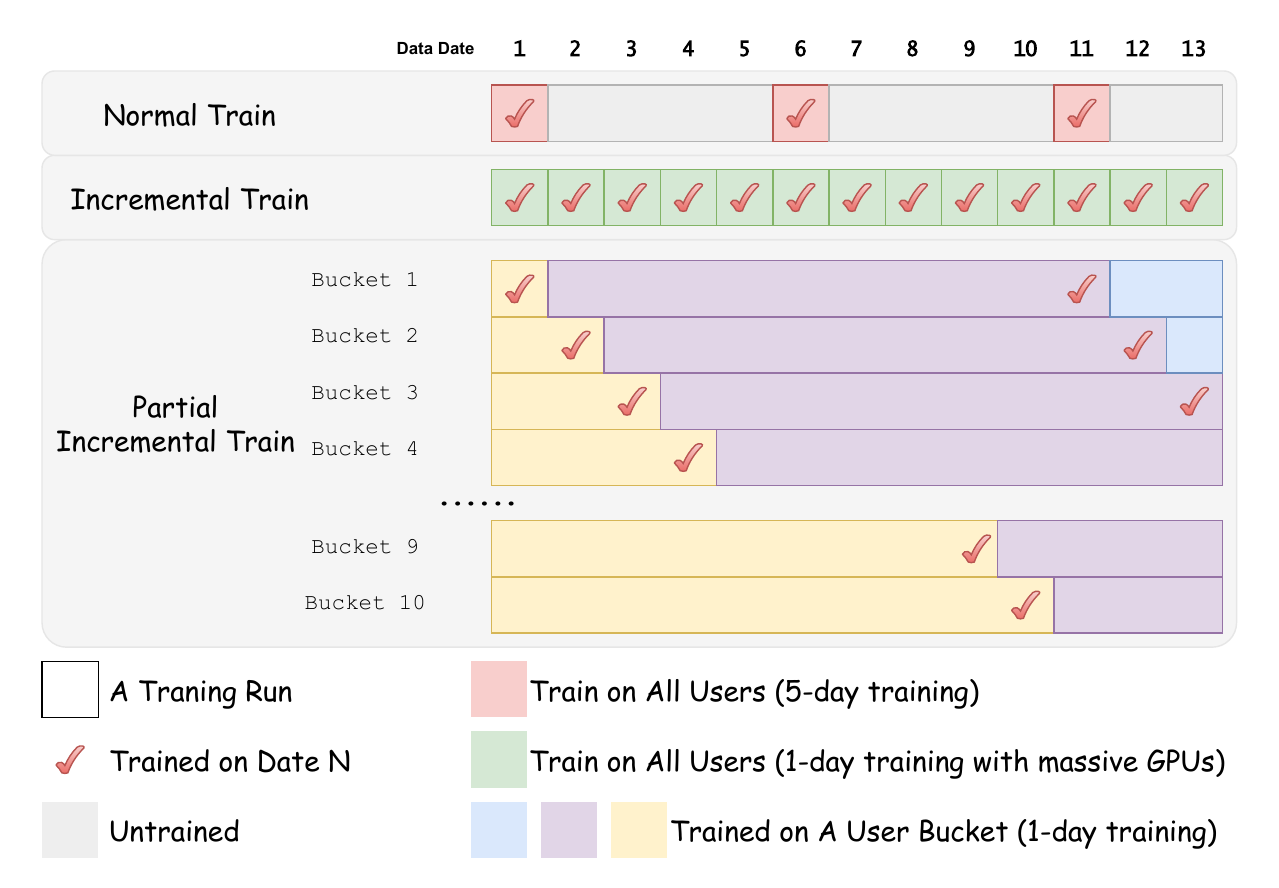}    }
\caption{Partial Incremental Training. In the figure, each date generates new user behavior data. All historical data are trained on the dates marked with a checkmark. Since PIT continuously iterates over each bucket, no data is ultimately lost. At each deployment, there is always one bucket containing the most recent data that has been trained.}
\vskip -0.15in
\label{incremental_Training}
\end{center}
\vskip -0.15in
\end{figure}


To address this issue, we propose Partial Incremental Train(PIT) as demonstrated in \cref{incremental_Training}. Specifically, all users are randomly partitioned into 10 buckets. During each incremental training phase, only the data from the most recent ten days in one bucket is used for training. Since only one-tenth of the full dataset is processed in each iteration, the training with a limited number of GPUs can be completed in less than a day, ensuring that the updated model is deployed within the next day. Moreover, because every user's data from each day is eventually covered across the iterations, there is no loss or underutilization of data. Given the sufficiently large data volume in each bucket, the data distribution within any given bucket remains representative of the overall distribution. Consequently, the proposed method achieves performance comparable to the ideal scenario.

\subsection{Inference}

During the inference phase, based on the current context and timestamp provided in the recommendation request from the client side, a new context token $c^{(K+1)}$ is appended to the end of the user sequence, after which the model generates the output token corresponding to the final position. This context token is then used to perform ANN search over the item pool to retrieve the top candidate items. Since the context token is trained with supervision from both in-session items and negative samples drawn from the item pool, it learns to capture the underlying semantics of potential retrieval items. Given the context embedding $h_{c^{(K+1)}}$, the retrieval score for item $i$ is:

\begin{equation}
    \text{Score}(c^{(K+1)}, i) = \langle \mathbf{h}'_{c^{(K+1)}}, \mathbf{e}_i \rangle
    = \langle \mathbf{h}'_{c^{(K+1)}}, \mathbf{e}_{i,\text{id}} + \mathbf{e}_{i,\text{side}} \rangle
\end{equation}

candidate items are retrieved via:

\begin{equation}
    \mathcal{R} = \arg\max_{i \in \mathcal{I}} \, \text{Score}(c^{(K+1)}, i).
\end{equation}

where $\mathcal{I}$ denotes the high-quality item pool.

%% file: Engineering_level_details.tex
\section{Engineering Level Details}

\subsection{Training Framework}

\begin{figure}[ht]
\vskip -0.1in
\begin{center}
\includegraphics[width=\columnwidth]{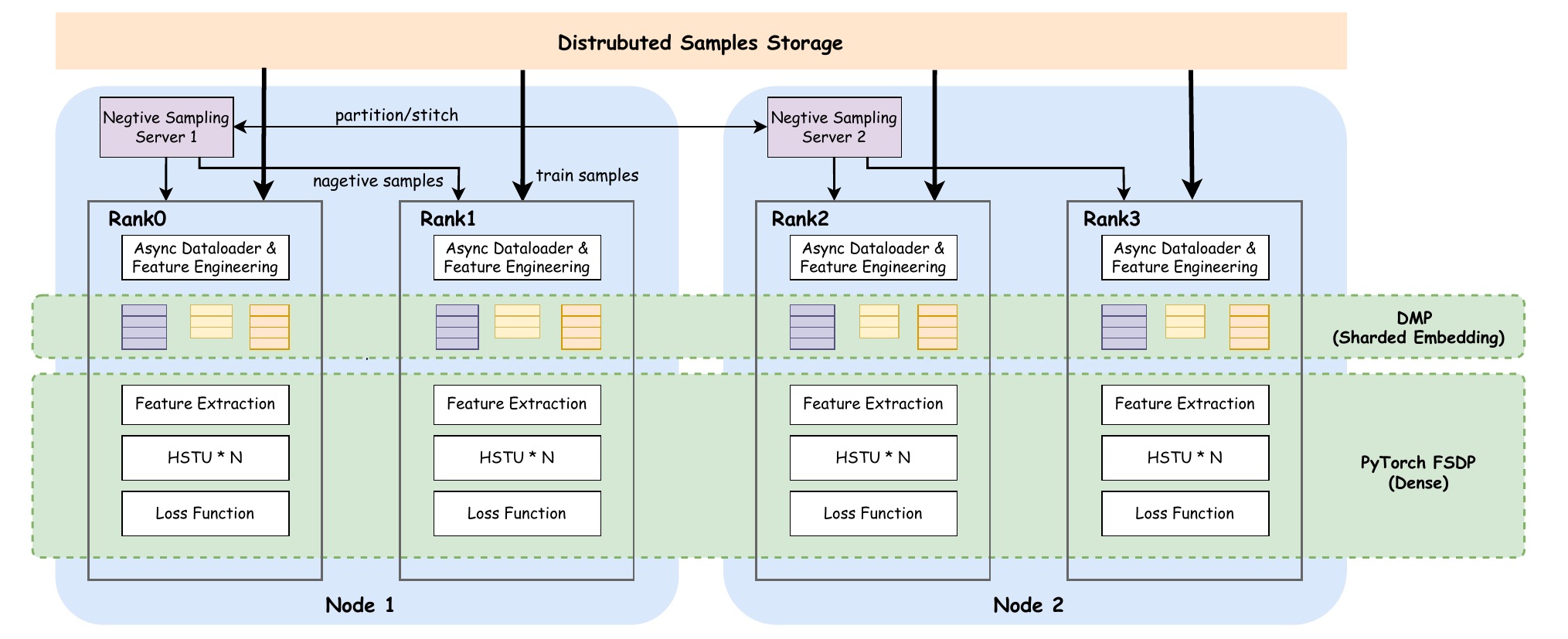}
\vskip -0.1in
\caption{Distributed Training Framework.}
\label{grs_training_topo}
\end{center}
\vskip -0.15in
\end{figure}
\begin{figure}[ht]
\vskip -0.2in
\begin{center}
\includegraphics[width=0.75\columnwidth]{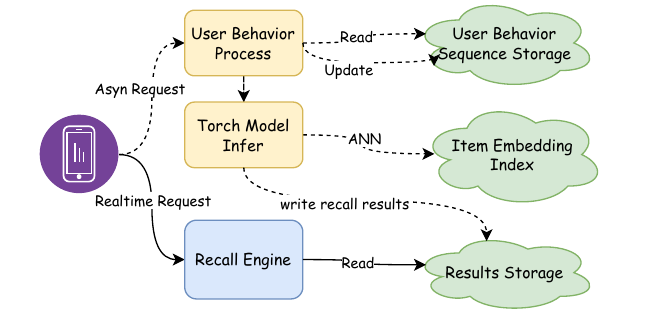}
\vskip -0.1in
\caption{Online Serving System.}
\label{online_deployment}
\end{center}
\vskip -0.15in
\end{figure}

A high-performance training framework in \cref{online_deployment} is built for our generative recommendation models in large-scale e-commerce systems. It addresses two challenges: (1) training models with 10B-scale sparse and billion-scale dense parameters on trillion-scale token corpora, and (2) enabling weighted random negative sampling over billion-scale product inventories for retrieval optimization. Built on TorchRec, our framework combines distributed systems and advanced training for production efficiency.





\begin{itemize}
\item \textbf{Distributed Negative Sampling} Shards billion-scale product catalogs across nodes, enabling low-latency distributed sampling.
\item \textbf{Asynchronous Data Loading} A CPU-based DataLoader uses prefetching/pipeline parallelism to hide I/O latency and maintain >90\% GPU utilization. It offloads data preparation tasks (e.g., feature engineering, negative sampling) to async CPU workers, leveraging CPU parallelism for non-GPU tasks in large-scale training.
\item \textbf{Sharded Embedding} TorchRec’s DMP partitions sparse embeddings row-wise across GPUs, reducing memory via AllToAll communication.
\item \textbf{FSDP} Partitions dense model parameters/optimizer states across devices for linear memory scaling, to support model training with larger parameter quantities.
\item \textbf{End-to-End Platform Capabilities} Supports daily retraining with fault-tolerant checkpointing, MLTracker metrics monitoring, TorchScript exporting, and consistency checks.
\end{itemize}

\subsection{Online Serving System}

To mitigate online inference compute/latency constraints, we develop a nearline retrieval framework that decouples user representation generation from request handling, as shown in Figure 5. Real-time Taobao user interactions (e.g., exposures, clicks) trigger two async services: (1) behavior sequence updates and (2) generative model inference. The latter constructs input token sequences from the latest user behavior data, executes PyTorch-based inference through the generative model to generate a user interest vector, and subsequently performs top-k item retrieval via an ANN search on a pre-built offline item embedding index. Results are cached as <userid, {topkItemList}> key-value pairs.

Online requests directly fetch precomputed recommendations from the cache, eliminating the need for real-time computation. This design reduces latency and shifts resource-intensive retrieval tasks to the nearline pipeline. Periodic interest vector updates balance freshness and computational efficiency, ensuring scalable, low-latency deployment.

%% file: experiment.tex
\section{Experiment}
\subsection{Setup}
\subsubsection{Dataset}
We conduct experiments on two large-scale datasets: 1) RecFlow \cite{recflow}, which contains rich contextual information from the Kuaishou. 2) TaoBao dataset, a real-world, industrial-scale dataset derived from traffic logs of TaoBao. \cref{tab:dataset-stats} summarizes the main statistics for these two datasets.
\begin{table}[ht]
\centering
\vskip -0.1in
\caption{The statistic of datasets.}
\vskip -0.1in
\label{tab:dataset-stats}
\begin{tabular}{lrrr}
\toprule
\textbf{Dataset} & \textbf{\#Interaction} & \textbf{\#User} & \textbf{\#Item} \\
\midrule
RecFlow & 30,255,736 & 30,968 & 3,242,884 \\
TaoBao Dataset & 1.2 trillion & 0.5 billion & 0.5 billion \\
\bottomrule
\end{tabular}
\vskip -0.1in
\end{table}
\subsubsection{Metrics} Hit Rate@K (HR@K) is adopted to evaluate the effectiveness of our proposed method. It is formally defined as:
\begin{equation}
\text{HR@}K = \frac{|R_u^{(K)} \cap G_u|}{|G_u|}, 
\end{equation}
where $R_u^{(K)}$ denotes the set of retrieved items ($|R_u^{(K)}|=K$) and $G_u$
denotes the set of ground truths.
\subsubsection{Baselines} To validate the superior performance of our model, we selected
multiple generative recommendation baseline models.
The models we selected are all non-LLM architectures that, like TBGRecall, utilize sparse IDs. HSTU, SASRec \cite{sasrec}, and BERT4Rec \cite{bert4rec} are sequential modeling-based recommendation models, while YouTubeDNN \cite{youtubednn} and FDSA \cite{fdsa} represent traditional recommendation approaches. The ONLINE model adopts Taobao’s production-level dual-tower baseline for recommendation retrieval. The ONLINE (DT) denotes that the model is updated through daily training over multiple years, ensuring sustained adaptation to long-term trends and dynamic user preferences. 
\subsubsection{Implementation details} Our model is trained using the Adam optimizer for dense parameters and Adagrad for sparse parameters, with an initial learning rate of $1 \times 10^{-3}$. 
The TBGRecall framework is deployed and optimized on the PPU 810E platform, an internal GPU architecture developed by Alibaba with a computational performance level approximately 60\% that of the NVIDIA A100.
During training,  we fix the HSTU parameters under a lightweight parameter configuration (Block = 4, Length = 5120, Dim = 512). For models involving the Mixture-of-Experts (MoE) architecture, we adopt the MoE design, which consists of 24 routed expert networks and 1 shared expert. In each forward pass, 2 experts are activated via a top-K selection mechanism. In addition, unless otherwise specified, all mentioned methods are trained from scratch with carefully optimized hyperparameters.



\subsection{Results}
We commpare our proposed TBGRecall with multiple generative recommendation baseline models on the RecFlow (RF) and TaoBao (TB) datasets, as illustrated in \Cref{baseline_table}.

On the RecFlow dataset, TBGRecall demonstrates superior performance across all HR@K metrics compared to state-of-the-art baselines (\Cref{baseline_table}). Notably, the gains are most significant in top-ranked results (HR@20–HR@500), achieving stronger improvements in high-priority item retrieval. The superior performance stems from the session-wise sequence construction and hierarchical contrastive objectives, which prioritizes high-value interactions in retrieval.

On Taobao's industrial-scale dataset, TBGRecall substantially outperforms all baselines across HR@K metrics, as shown in \Cref{baseline_table}. Notably, TBGRecall achieves superior HR performance compared to both the \textbf{Online} baseline trained from scratch and the production-grade daily-trained variant \textbf{Online (DT)}.

\begin{table}[ht]
\begin{sc}
\label{TBRecall vs variants}
\begin{small}
\vskip -0.1in
\caption{Comparison of Different Model Variants.}
\vskip -0.1in
\centering
\begin{tabular}{lrrrr}
\toprule
\textbf{Model} & \textbf{HR@20} & \textbf{HR@500} & \textbf{HR@1000} & \textbf{HR@4000} \\
\midrule
TBGRecall       & \textbf{1.53\%}           & \textbf{11.81\%}           & \textbf{16.62\%}            & \textbf{29.45\%}           \\
w/o TSN         & 1.51\%           & 11.67\%           & 16.42\%            & 28.88\%           \\
w/o MSP         & 1.35\%           & 10.59\%           & 15.08\%            & 27.28\%           \\
w/o MoE         & 1.46\%           & 11.41\%           & 16.08\%            & 28.47\%           \\
w/o sw-RoPE        & 1.32\%           & 10.67\%           & 15.20\%            & 27.36\%           \\
\bottomrule
\end{tabular}
\vskip -0.1in
\end{small}
\end{sc}
\end{table}

\subsection{Ablation Study}
We conducted comprehensive ablation experiments on key components including TSN, MSP, MoE\cite{deepseekv3}, and RoPE\cite{rope} to assess their individual contributions to model performance. Results confirm that each module independently enhances retrieval accuracy, with the full framework achieving the highest overall effectiveness. MSP contributes the most significant performance improvement, primarily because it increases the amount of information encoded within a single session, enabling the model to better understand the contextual semantics and avoid falling into local decision-making patterns during training. 
Both session-wise RoPE (Rotary Position Embedding) and MoE (Mixture of Experts) are widely adopted techniques in current state-of-the-art large language models. By incorporating them into our architecture, we achieve considerable gains in model performance. Additionally, the TSN (Token-Specific Network) module enables context tokens and item tokens to align more effectively and efficiently, further enhancing the overall performance of the model. These findings validate the architectural soundness and practical value of the proposed components in advancing generative retrieval systems for e-commerce recommendation.

\subsection{Partial Incremental Training}

\begin{table}[H]
\begin{small}
\vskip -0.05in
\caption{Partial Incremental Training. \textit{Incre}: Daily Incremental Training, \textit{SL}: Stochastic Length in HSTU, \textit{PIT}: Partial Incremental Training (Ours)}
\vskip -0.1in
\centering
    \begin{tabular}{l|cccc}
\toprule
        ~ &\textbf{Duration}&\textbf{GPUs}&\textbf{Latest Data Used}&\textbf{HR@4000}\\
\midrule
        \textbf{Normal} & 5d & 128 & N & 23.76\%\\ \hline
        \textbf{Incre} & 16h & 1280 & Y & \textbf{29.76\%}\\ \hline
        \textbf{SL} & 48h & 128 & N & 26.50\%\\ \hline
        \textbf{PIT} & 11h & 128 & Y & \underline{29.45\%}\\
\bottomrule
\end{tabular}
\vskip -0.1in
\label{incremental_training_table}
\end{small}
\end{table}

\begin{table*}[!ht]
\begin{sc}
\caption{General Performance on two datasets.}
\vskip -0.1in
    \centering
    \begin{tabular}{l|cccccc|cccccc}
    \toprule
        \textbf{MODEL} & \multicolumn{6}{c|}{\textbf{RecFlow}} & \multicolumn{6}{c}{\textbf{TaoBao}} \\
        \cline{1-13}
        \textbf{HR}&  {\textbf{@20}} &  {\textbf{@100}} &  {\textbf{@500}} &  {\textbf{@1000}} &  {\textbf{@2000}} &  {\textbf{@4000}} & {\textbf{@20}} &  {\textbf{@100}} &  {\textbf{@500}} &  {\textbf{@1000}} &  {\textbf{@2000}} &  {\textbf{@4000}} \\
        \midrule
\textbf{YoutubeDNN} & 0.01\% & 0.08\% & 0.43\% & 1.02\% & 2.37\% & 3.28\% & 0.47\% & 1.29\% & 3.28\%  & 4.75\%  & 6.68\%  & 9.22\%  \\
\textbf{SASRec}     & 0.04\% & 0.19\% & 1.00\% & 1.96\% & 3.66\% & 6.53\% & 0.50\% & 1.65\% & 4.47\%  & 6.40\%  & 8.79\%  & 11.70\% \\
\textbf{Bert4Rec}   & 0.06\% & 0.24\% & 1.01\% & 1.87\% & 3.43\% & 6.09\% & 0.11\% & 0.44\% & 1.53\%  & 2.50\%  & 3.90\%  & 5.96\%  \\
\textbf{FDSA}       & 0.02\% & 0.13\% & 0.54\% & 1.41\% & 3.90\% & 4.32\% & 0.52\% & 1.47\% & 3.70\%  & 5.30\%  & 7.44\%  & 10.14\% \\
\textbf{HSTU}       & \underline{0.06\%} & \underline{0.30\%} & \underline{1.21\%} & \underline{2.19\%} & \underline{4.06\%} & \underline{7.38\%} & 0.54\% & 1.78\% & 4.88\%  & 7.01\%  & 9.75\%  & 13.11\% \\
\textbf{Online} & -    & -    & -    & -    & -    & -   &  1.03\% & 3.53\% & 9.37\% & 13.36\% & 18.08\% & 23.57\%  \\
\textbf{Online(DT)}  & -    & -    & -    & -    & -    & -  & \underline{1.26\%} & \underline{4.28\%} & \underline{11.03\%} & \underline{15.44\%} & \underline{20.62\%} & \underline{26.45\%}   \\
\textbf{TBGRecall}  & \textbf{0.26\%} & \textbf{0.83\%} &  \textbf{2.45\%} & \textbf{3.85\%} & \textbf{5.61\%} & \textbf{8.20\%} & \textbf{1.53\%}  & \textbf{4.65\%}  & \textbf{11.81\%} & \textbf{16.62\%} & \textbf{22.57\%} & \textbf{29.45\%}  \\
    \bottomrule
\end{tabular}
\vskip -0.1in
\label{baseline_table}
\end{sc}
\end{table*}


\Cref{incremental_training_table} presents the experimental results under different training strategies, where Train Duration denotes the time required to train one day's worth of user data, and Latest Data Used indicates whether the most recent day's data was included in the training before model deployment. Due to a deployment delay of ten days, normal training does not yield superior performance. In contrast, when using a large number of GPUs to perform daily incremental training, there is a significant improvement in model performance. Using Stochastic Length in HSTU, with $\alpha = 1.6$, can reduce training time by 70\% but fails to promptly utilize the latest data,  and causes 9\% performance drop. However, our partial incremental training approach still maintains competitive performance while utilizing only one-tenth of the GPU resources. This strongly demonstrates that our partial incremental training method is highly suitable for training large-scale generative models.

\subsection{Scaling Law}

\begin{figure}[ht]
\vskip -0.15in
\begin{center}
\centerline{\includegraphics[width=\columnwidth]{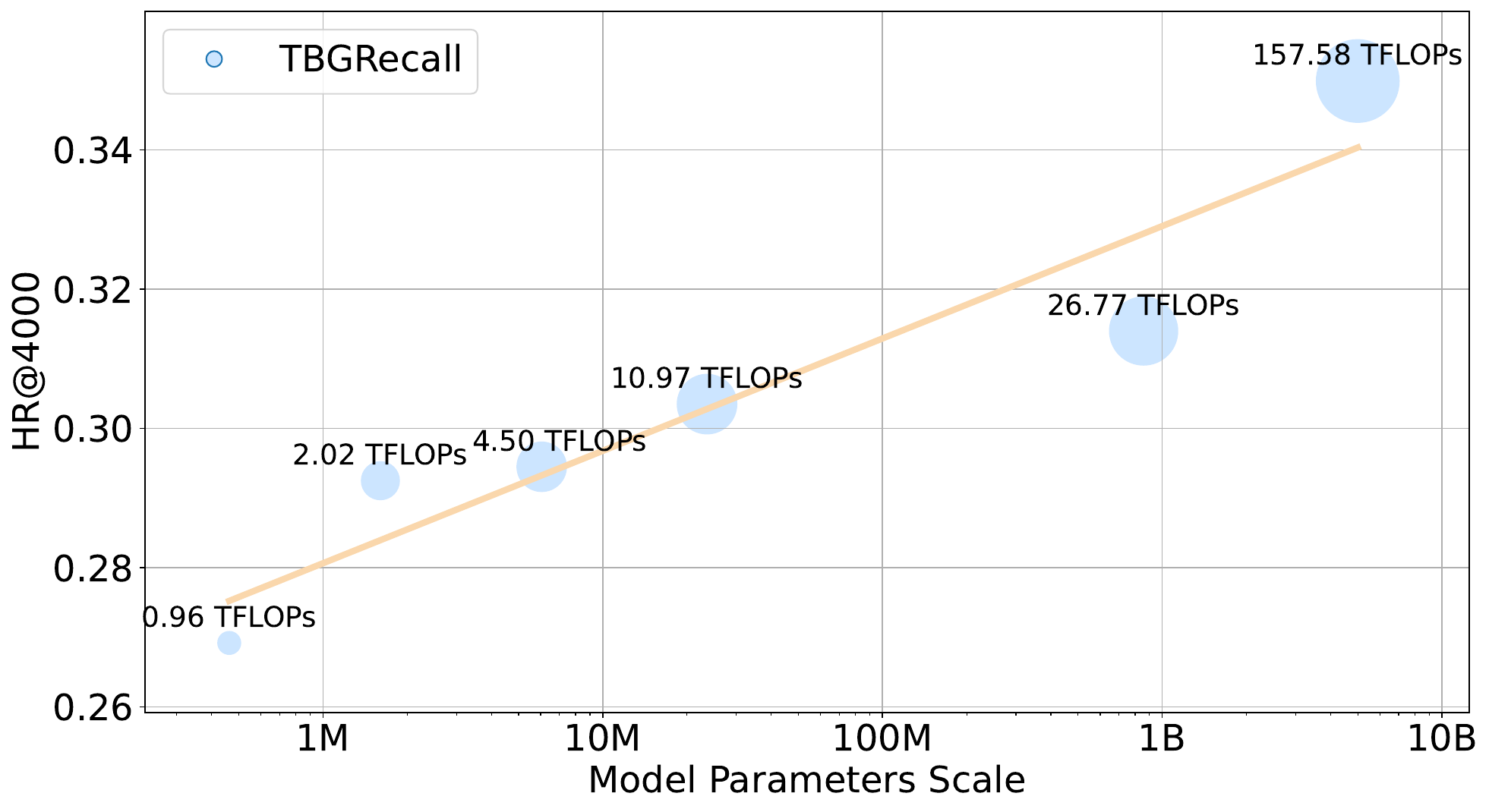}    }
\vskip -0.1in
\caption{Scaling Law.}
\label{Scaling_Law}
\end{center}
\vskip -0.2in
\end{figure}

We systematically evaluated the relationship between model capacity, computational resources, and retrieval performance. In \cref{Scaling_Law}, we scale up the model size by gradually increasing the hidden layer dimensions (128, 256, 512, and 1024, respectively). The two rightmost data points correspond to models utilizing sparse MoE (Mixture of Experts), with the latter one employing 24 hstu blocks. As the number of parameters increases, the computational FLOPS also grow exponentially. By taking the logarithm of the horizontal axis, we observe a linear relationship between the logarithm of model parameters and model performance. Experimental results demonstrate that model accuracy exhibits a scaling law with respect to both parameter count and training computation, indicating consistent performance gains as model scale increases. Notably, the proposed framework maintains efficient scaling behavior even at large-scale industrial settings, confirming the feasibility and effectiveness of deploying highly parameterized generative models in real-world recommendation systems.



\subsection{Online A/B Test}

\begin{table}[ht]
\begin{sc}
\begin{small}
\vskip -0.1in
\caption{Online A/B Testing Results.}
\centering
\begin{tabular}{ccc}
\toprule
\textbf{PVR} & \textbf{Transaction Count} & \textbf{Transaction Amount} \\
\midrule
23.94\% & +0.60\% & +2.16\% \\
\bottomrule
\end{tabular}
\vskip -0.1in
\label{online_ab_test}
\end{small}
\end{sc}
\end{table}

We conduct online A/B testing on the "Guess You Like" section of Taobao's homepage, which generates hundreds of millions of daily exposures. A new retrieval strategy based on the TBGRecall paradigm is added to the retrieval stage. Experiments run for 7 days on 5\% randomized user traffic. 

As shown in \Cref{online_ab_test}, our strategy captures 23.94\% of total exposures among all deployed retrieval methods. This highlights its competitiveness in a production environment with multiple concurrent strategies. Notably, statistically significant improvements are observed: +0.60\% in transaction count and +2.16\% in transaction amount. These gains validate the method’s effectiveness in large-scale industrial deployment. It balances high traffic allocation with stable business impact, making it suitable for real-world recommendation systems requiring scalability and performance.

%% file: conclusion.tex
\section{Conclusion}

This work introduces a paradigm innovation for generative recommendation systems through Next Session Prediction (NSP), resolving inherent limitations of autoregressive generation in retrieval scenarios. The proposed TBGRecall model redefines sequence modeling via session-wise autoregression, eliminating intra-session sequential dependencies while capturing cross-session behavioral patterns through context token optimization. 
The framework achieves improved predictive accuracy and computational efficiency, with superior performance demonstrated over baselines in large-scale e-commerce settings. 
The establishment of scaling laws in sparse ID-based architectures and validation of data recency principles further provide actionable insights for industrial deployment. By bridging generative modeling with practical retrieval infrastructure, this work advances the feasibility of session-aware sequential recommendation models, offering both robust industrial solutions and directions for future research.